\documentclass[12pt]{article}
\usepackage{epsfig}

\usepackage{graphicx}
\usepackage{stackengine}
\usepackage{cancel}
\usepackage{amssymb,amsmath}
\def\harr#1#2{\smash{\mathop{\hbox to .3in{\rightarrowfill}}
 \limits^{\scriptstyle#1}_{\scriptstyle#2}}}

\def\s2{\frac{1}{\sqrt2}}

\def\be{\begin{equation}}
\def\ee{\end{equation}}
\def\beqa{\begin{eqnarray}}
\def\eeqa{\end{eqnarray}}

\def\Dsl{\,\raise.15ex\hbox{/}\mkern-13.5mu D} %can be subscripted
\def\d3{d^3}

\newcommand{\dif}{\mathrm{d}}
\newcommand{\im}{\mathrm{i}}

\begin{document}
%\rightline{\tt hep-th/yymmnnn }

\begin{center}
\Large{\bf On an integrable deformation of Kapustin-Witten systems}
\vspace{0.5cm}

\large  S.A.H. Cardona$^a$\footnote{e-mail address: {\tt sholguin@im.unam.mx}},
           H. Garc\'{\i}a-Compe\'an$^{b,c}$\footnote{e-mail address: {\tt compean@fis.cinvestav.mx}},
           A. Mart\'{\i}nez-Merino$^{c,d}$\footnote{e-mail address: {\tt a.merino@fisica.ugto.mx}}\\

\vspace{0.5cm}

{\small \em $^a$Investigador C\'atedra CONACyT, Instituto de Matem\'aticas, Universidad Nacional Aut\'onoma de M\'exico, Unidad Oaxaca}\\
{\small \em Le\'on No. 2, Col. centro, C.P. 68000, Oaxaca de Ju\'arez, Oax., M\'exico.}\\
\vspace{0.5cm}
{\small \em $^b$Departamento de F\'{\i}sica, Centro de Investigaci\'on y de Estudios Avanzados del IPN}\\
{\small\em P.O. Box 14-740, CP. 07000, M\'exico D.F., M\'exico.}\\
\vspace{0.5cm}
{\small \em $^c$Departamento de F\'{\i}sica, Divisi\'on de Ciencias e Ingenier\'{\i}a}\\
{\small\em Universidad de Guanajuato, Campus Le\'on}\\
{\small\em Loma del Bosque No. 103, Frac. Lomas del Campestre, Le\'on, Gto., M\'exico.}\\
\vspace{0.5cm}
{\small \em $^d$Investigador C\'atedra CONACyT, Facultad de Ciencias en F\'isica y Matem\'aticas}\\
{\small\em Universidad Aut\'onoma de Chiapas}\\
{\small\em Carretera Emiliano Zapata Km. 08, Rancho San Francisco Col. Ter\'an}\\
{\small\em Ciudad Universitaria, Tuxtla Guti\'errez, Chiapas, M\'exico.}\\
\vspace*{0.5cm}
\end{center}

\begin{abstract}
In this article we study an integrable deformation of the
Kapustin-Witten equations. Using the Weyl-Wigner-Moyal-Groenewold
description an integrable $\star$-deformation of a Kapustin-Witten
system is obtained. Starting from known solutions of the original
equations, some solutions to these deformed equations are obtained.

\vskip .3truecm


\noindent\leftline{November 1, 2017}
%\noindent{\it MS Classification}: 53C07, 53C55, 32C15.
\end{abstract}

\bigskip
% \leftline{August 06, 2017} Para poner fecha ''a mano''
\newpage
%%%%%%%%%%%%%%%%%%%%%%%%%%%%%%%%%%%%%%%%%%%%%%%%%%%%%%%%%%%%%%%%%%%%%%%%%%%%%%%%%%%%%%
%%%%%%%%%%%%%%%%%%%%%%%%%%%%%%%%%%%%%%%%%%%%%%%%%%%%%%%%%%%%%%%%%%%%%%%%%%%%%%%%%%%%%%
%%%%%%%%%%%%%%%%%%%%%%%%%%%%%%%%%%%%%%%%%%%%%%%%%%%%%%%%%%%%%%%%%%%%%%%%%%%%%%%%%%%%%%
\section{Introduction}

In a celebrated work, Kapustin and Witten \cite{Kapustin:2006pk}
described the geometric Langlands program (GLP) in terms of a
compactification on a Riemann surface of a certain twisted version
of the ${\cal N}=4$ superymmetric Yang-Mills theory (SYM) in four
dimensions. In such paper, the authors introduced a set of equations
after imposing a BRST-like preservation conditions on a twisted
version of ${\cal N}=4$ SYM theory in four dimensions; these
equations are now known as the Kapustin-Witten (KW) equations and
have been the subject of an intensive work in the last decade in
physics as well as in mathematics. In particular, a relation of KW
equations with knot theory is also described by Witten in
\cite{Witten:2011zz}, where the author describes an approach to
Khovanov homology using gauge theory; in that context, the KW
equations appear as a localization condition of the ${\cal N}=4$ SYM
theory in four dimensions (see \cite{khovanov} for a review on this
topic). The KW equations are also closed related to another set of
equations, recently introduced by Ward \cite{Ward:2016ygr} and
usually called the $(2k)$-Hitchin equations; it is important to
mention that these  equations are a natural generalization of
another set of equations introduced by Hitchin \cite{Hitchin:1987}
in a pionnering work in complex geometry; indeed, the article of
Hitchin is the origin of the notion of Higgs bundle in mathematics,
a notion that plays an important role in the physical interpretation
of the GLP developed by Kapustin and Witten.

In this section we fix the notation and review some preliminar
notions that will be used in the article. In order to start, let
$\mathfrak{g}$ be a Lie algebra of a Lie group $G$ and let $M$ be a
riemannian 4-manifold, with riemannian metric $g$. Let $\phi$ and
$A$ be two $\mathfrak{g}$-valued 1-forms, where $\phi$ is considered
as a Higgs field and  $A$ is a gauge potential with covariant
derivative $D=d+A$. As it is well known, the gauge field of $A$ is a
$\mathfrak{g}$-valued 2-form given by $F=D\wedge D=dA + A\wedge A$.
With this data, Kapustin and Witten introduced in
\cite{Kapustin:2006pk} the following set of equations:
\begin{equation}
(F-\phi\wedge\phi +t\,D\phi)^{+}=0\,,\quad\quad (F-\phi\wedge\phi - t^{-1}D\phi)^{-}=0\,,\quad\quad D^{*}\phi=0\,.  \label{KW-eqs.}
\end{equation}
Here the superscript $^{\pm}$ stands for the self-dual and anti-self-dual part, $t$ is  a complex parameter and $D^{*}=*D*$, where $*$ is the Hodge operator on $M$ associated with
$g$. Strictly speaking (\ref{KW-eqs.}) is not a system of equations, but a family of equations parametrized by $t$ and certain values of $t$  give rise to equations of interest, e.g., with $t=\pm i$ we obtain a particular case which plays an important role in the GLP approach developed in \cite{Kapustin:2006pk}; when $t=1$ the equations take the form
\begin{equation}
F-\phi\wedge\phi +*D\phi =0\,,\quad\quad D*\phi=0\,, \label{5Branes} % Fivebranes and knots, (3.6) pag. 40. ''t'' es complejo (raiz de (\bar\tau)/\tau)
\end{equation}
which are equations\footnote{To be more precise, in that case there exists a dual parameter $t^{\vee}$ of $t$, and the dual condition $t^{\vee}=1$ gives such a set of equations with respect to the dual group $G^{\vee}$.} considered in \cite{Witten:2011zz}; see also \cite{Mazzeo&Witten}. Notice that with $t=1$ and assuming also that $F-\phi\wedge\phi$ is self-dual and $D\phi$ is anti-self-dual, eqs. (\ref{KW-eqs.}) become
\begin{equation}
(F-\phi\wedge\phi)^{+}=0\,,\quad\quad (D\phi)^{-}=0\,,\quad\quad D^{*}\phi=0\,.  \label{t=1 KW-eqs.}
\end{equation}
This form of KW equations have been considered recently by Gagliardo and Uhlenbeck \cite{Gagliardo-Uhlenbeck:2012}, Dunajski and Hoegner \cite{Dunajski:2011sx} and Ward \cite{Ward:2016ygr}, in particular in the last two references the equations (\ref{t=1 KW-eqs.}) are also called the {\it non-abelian Seiberg-Witten equations}. From now on, we will refer to (\ref{KW-eqs.}) and (\ref{t=1 KW-eqs.}) as the KW equations and the non-abelian Seiberg-Witten equations, respectively. At this point, it is important to mention that some solutions to (\ref{t=1 KW-eqs.}) are already proposed by Dunajski and Hoegner in \cite{Dunajski:2011sx}, these solutions will be of crucial importance for the purposes of the present paper; indeed, we are going to use such solutions to find solutions of the deformed equations.

Now, KW equations are close related to self-duality in higher
dimensions and to the dimensional reduction procedure. Moreover, as
we said before such equations appear as equations of motion of a
twisted version of ${\cal N}=4$ SYM in four dimensions, which in
turn arises as a dimensional reduction to four dimensions of the
${\cal N}=1$ SYM theory in ten dimensions \cite{Kapustin:2006pk}.
Also, in Ref. \cite{Ward:2016ygr} it was shown that KW equations
arises as a four dimensional reduction of the self-dual YM equations
in eight dimensions. Moreover, the five dimensional extension of KW
equations proposed in \cite{Witten:2011zz} for a particular case,
leads to a set of equations which had already been obtained for
general five-manifolds by Haydys in \cite{haydys}. They are known as
Haydys-Witten equations and later on it was found by Cherkis
\cite{Cherkis:2014xua} that these equations can be obtained via
dimensional reduction from the instanton equation on Spin(7)
eight-manifolds and also from the seven-dimensional reductions for
any $G_2$ holonomy manifold. All this shows that dimensional
reduction and integrability are strongly related to KW equations and
play an important role in the undestanding of all story about these
equations. In fact, since many years ago it is well known that
Seiberg-Witten equations have the structure of integrable systems
\cite{Donagi:1995cf} (see \cite{Olshanetsky:2009zv} for an
overview). Moreover, the Hitchin systems \cite{Hitchin:1987} are
very well known examples of integrable systems that have been
studied even in a quantized way \cite{drinfeld}. It is also known
that self-dual systems are integrable systems and according the Ward
conjecture, all integrable systems come from four-dimensional
self-dual YM or self-dual gravity equations \cite{Wardconjecture}.

In the present paper we perform an integrable deformation of the KW
equations via the Weyl-Wigner-Moyal-Groenewold (WWMG) formalism of
deformation quantization (for a recent overview see
\cite{Zachos:2001ux}). As it is well known, this procedeure does not
spoil the integrability of the former equations
\cite{Strachan:1992em,Takasaki:1992jf,Takasaki:1993my,Strachan:1996gx,Formanski:2004dd,Formanski:2005wt,dgs}.
For a more recent review containing several of these results the
reader would like to consult \cite{Mason:1991rf,Dunajski:2010zz}.
Thus, in the present paper we will find integrable deformations of
KW equations. In order to find solutions for these deformed
equations, we make use of the WWMG correspondence between
$\mathfrak{su}(2)$-valued operators acting on a certain Hilbert
space and functions defined in a symplectic surface satisfying
similar relations but under the Moyal bracket. It is important to
mention that  this correspondence has been already explored before
in other contexts, indeed in
\cite{Plebanski:1995jm,GarciaCompean:1996np,GarciaCompean:2009cg} it
was employed to study some integrable deformations of the principal
chiral model, the Nahm equations and seven-dimensional reductions of
the self-dual in eight-dimensions respectively.

This paper is organized as follows: in Section 2 we start with a
general overview of the KW equations and we perform the deformation
of these equations. In Section 3 we analyze the non-abelian
Seiberg-Witten equations, since they have a close relation with our
former system KW. In that section we perform the integrable
deformation of this system under the same WWMG formalism and found
some solutions to it. In Section 4 we close the paper with some
final comments. Since we consider this subject a very important
matter, we left to the appendix a unification in terms of language
of the Hitchin equations in the believing that it would come in
handy to physicists and mathematicians alike.

%%%%%%%%%%%%%%%%%%%%%%%%%%%%%%%%%%%%%%%%%%%%%%%%
%%%%%%%%%%%%%%%%%%%%%%%%%%%%%%%%%%%%%%%%%%%%%%%%

\section{Overview on Kapustin-Witten equations}

We begin this section by reviewing the general KW equations. As we
mentioned in the introduction, these equations arise as the
equations of motion of a topological twisting of ${\cal N}=4$ SYM in
four dimensions  \cite{Kapustin:2006pk}, or also as a set of
equations obtained to localize knots in 3-dimensional space
\cite{Witten:2011zz}. By defining
\begin{equation}
\mathcal{V}^+ = (F - \phi \wedge \phi + t\,D \phi )^+\,, \quad\quad
\mathcal{V}^- = (F - \phi \wedge \phi - t^{-1} D \phi )^-\,,
\quad\quad \mathcal{V}^0 = D_{\mu} \phi^\mu, \nonumber
\end{equation}
where $^{\pm}$ stands for the self-dual and anti-self-dual
projections of the 2-forms between parenthesis, the KW equations
(\ref{KW-eqs.}) can be written simply as
\begin{equation}
\mathcal{V}^+ =\mathcal{V}^- = \mathcal{V}^0 =0\,.   \label{LocEqs}
\end{equation}
As it is explained in \cite{Kapustin:2006pk}, these equations are
obtained as equations of motion from the action
\begin{equation}
S = - \int_{M} \dif^4 x \sqrt{g}\; \mathrm{Tr} \left[\frac{t^{-1}}{t + t^{-1}} \mathcal{V}_{\mu \nu}^+
\mathcal{V}^{+\mu \nu} + \frac{t}{t + t^{-1}} \mathcal{V}_{\mu \nu}^{-} \mathcal{V}^{-\mu \nu} +
(\mathcal{V}^0)^2 \right],
\end{equation}
where $\sqrt g$ is an abbreviation for the square root of the determinant of the metric $g$ of $M$ and ${\rm Tr}$ is the usual trace. The fields $\phi$ and $F$ are real valued if the parameter $t$ is also real, in the present paper we are interested mainly in the case of real $t$. If we consider $x^\mu$, $\mu = 0, \dots, 3$, as the coordinates of the four-dimensional manifold $M$ and if we assume it has no boundary (see \cite{Kapustin:2006pk} for details) the action can be rewritten as
\begin{eqnarray}
S &=& -\int \dif^4 x \sqrt{g}\; \mathrm{Tr} \left[\frac{1}{2}
F_{\mu \nu}F^{\mu \nu} + D_\mu \phi_\nu D^\mu \phi^\nu + R_{\mu \nu} \phi^\mu \phi^\nu + \frac{1}{2}
[\phi_\mu, \phi_\nu][\phi^\mu, \phi^\nu] \right] \nonumber \\
& & + \,\frac{t - t^{-1}}{t + t^{-1}} \int_M \mathrm{Tr}\,(F \wedge F)\,,  \label{action}
\end{eqnarray}
where $R_{\mu \nu}$ is the Ricci tensor of $M$. In this form, the action is given as a sum of two terms in which the dependence on $t$ is reduced to the second one, which is indeed a topological term.

On the other hand, in \cite{Witten:2011zz} Witten find solutions to
eqs. (\ref{5Branes}) using the ansatz $A_0 = \phi_3 = 0$, in that
case, he showed that the resulting equations can be written nicely
in the form
\begin{eqnarray}
[\mathcal{D}_i, \mathcal{D}_j] &=& 0, \qquad i, j = 1, \dots, 3. \nonumber \\
\sum_{i = 1}^3 [\mathcal{D}_i, \mathcal{D}_i^\dagger] &=& 0, \label{HYM}
\end{eqnarray}
where the operators $\mathcal{D}_{i}$ are defined as follows:
\begin{equation}
\mathcal{D}_1 = {\partial}_{1} + i {\partial}_{2} + [A_1 + i A_2, \cdot\; ], \quad\quad \mathcal{D}_2 = {\partial}_{3} + [A_3 - i \phi_0, \cdot\; ], \quad\quad \mathcal{D}_3 = [\phi_1 - i \phi_2, \cdot\; ]. \nonumber
\end{equation}
Now, an important fact to note here is that equations of the same form than (\ref{HYM}) appear in the context of complex geometry and are usually called the Hermite-Yang-Mills equations (see \cite{Witten:2011zz} for more details). In that context, these equations are defined for a holomorphic vector bundle and the Hitchin-Kobayashi correspondence\footnote{This correspondence, also called the Uhlenbeck-Yau-Donaldson-Simpson theorem, plays a fundamental rol in complex geometry; in fact, it establishes an equivalence between the algebraic notion of Mumford stability and the differential notion of Hermite-Yang-Mills metric.} says that solutions to these equations exists, if and only if, the holomorphic bundle is poly-stable, i.e., it is a direct sum of stable bundles with the same slope (see \cite{Simpson} for details).

\indent In the present paper, we are only focus in gauge configurations with the Lie algebra being $\mathfrak{su}(2)$. Let $t_a$, $a = 1, \dots, 3$ be its generators in an anti-hermitian
representation. Thus, as it is shown in \cite{Witten:2011zz}, after complexifying the group and choosing some holomorphic gauge, the ansatz for solving (\ref{HYM}) is
\begin{eqnarray}
A_1 + i A_2 &=& -\frac{(\partial_1 + i \partial_2)v}{2} \left(
\begin{array}{cc}
1 & 0 \\
0 & -1
\end{array} \right), \nonumber \\
\phi_0 &=& -\frac{i \partial_3 v}{2} \left(
\begin{array}{cc}
1 & 0 \\
0 & -1
\end{array} \right), \nonumber \\
\varphi &=& z^{\mathfrak{r}} e^v \left(
\begin{array}{cc}
0 & 1 \\
0 & 0
\end{array} \right),
\end{eqnarray}

\noindent for an unknown function $v$ and where $\mathfrak{r}$ is a
parameter linked to the spin representation for the complexification
of the gauge group $\mathrm{SU}(2)$. The field strength associated
with this gauge field is
\begin{equation}
F_{12} = \frac{i (\partial_1^2 + \partial_2^2)v}{2} \left(
\begin{array}{cc}
1 & 0 \\
0 & -1
\end{array} \right).
\end{equation}

In order to $v$ solve equations (\ref{HYM}) it must be a solution to
\begin{equation}
-\left( \frac{\partial^2}{\partial (x^1)^2} + \frac{\partial^2}{\partial (x^2)^2} + \frac{\partial^2}{\partial y^2} \right) v + |z|^{2 \mathfrak{r}} e^{2v} = 0, \label{Eqforv}
\end{equation}
which come from the second line of (\ref{HYM}), where the definition
$y = x^3$ and $z = x^1 + i x^2$ was made. As it is known, the exact
solution to this equation is given by
\begin{equation}
v = -\mathfrak{r} \log |z| - \log y. \nonumber
\end{equation}

\noindent Depending on the class of the solution we are interested,
which represents the position where we are inserting the `t Hooft
operator in the dual description to the D3-NS5 system, $v$ can be
redefined in order to give the desired behavior. At the moment, we
want to emphasize that the solutions we just found are suitable of
being deformed. At this point, let us review the WWMG formalism in
order to apply it to the case considered here.

%%%%%%%%%%%%%%%%%%%%%%%%%%%%%%%%%%%%%%%%

\subsection{Deforming Kapustin-Witten equations}

\noindent In order to apply the WWMG formalism \cite{Zachos:2001ux},
we promote the fields $A$ and $\phi$ to $\mathfrak{su}(2)$
operator-valued forms acting over a Hilbert space $\mathcal{H} =
L^2(\mathbb{R})$. Let us choose $| \psi_n \rangle$, $n = 0, 1,
\dots$ an orthonormal basis of $\mathcal{H}$. As it is well known,
we have the closure relations
\begin{equation}
\langle \psi_n | \psi_m \rangle = \delta_{nm}, \qquad \sum_n |\psi_n
\rangle \langle \psi_n | = \widehat{I},     \nonumber
\end{equation}
where $\widehat{I}$ denotes the identity operator in $\mathcal{H}$.
We carry out the above-mentioned identification by the
correspondence: $A_i \rightarrow \widehat{A}_i \in
\mathfrak{M}\otimes \widehat{\mathcal{U}}$, and $\phi_i
\rightarrow\widehat{\phi}_i \in \mathfrak{M} \otimes
\widehat{\mathcal{U}}$, with $\widehat{\mathcal{U}}$ the Lie algebra
of anti-self-dual operators acting on $\mathcal{H}$. Also we change
the usual Lie algebra brackets by the corresponding commutator
$[\cdot, \cdot]$.

This deformation procedure rely on the parameter $\hbar$, and when
the limit $\hbar \to 0$ is taken we recover the (undeformed)
original system. In order to do this, we perform one further
redefinition of the fields in terms of $\hbar$,
\begin{equation}
\widehat{\mathcal{A}}_i = \im \hbar \widehat{A}_i, \qquad
\widehat{\Phi}_i = \im \hbar \widehat{\phi}_i.     \nonumber
\end{equation}

\indent Let $\mathcal{B}$ and $C^{\infty}(\Sigma, \mathbb{R})$
denote the  set of self-adjoint linear operators acting on the
Hilbert space $\mathcal{H} = L^2(\mathbb{R})$ and the space of
infinite differentiable real functions defined on the
two-dimensional phase space $\Sigma$ with coordinates $p,q$,
respectively. In general, we define the Weyl correspondence
$\mathcal{W}^{-1}: \mathcal{B} \to C^\infty (\Sigma,\mathbb{R})$ by
\begin{equation}
\mathcal{A}_i (\vec{x}, p, q; \hbar) \equiv \mathcal{W}^{-1}
(\widehat{\mathcal{A}}_i) := \int_{-\infty}^{\infty} \left\langle q
- \frac{1}{2}\xi \Big| \widehat{\mathcal{A}}_i (\vec{x}) \Big| q +
\frac{1}{2}\xi \right\rangle e^{\frac{\im}{\hbar} \xi \cdot p}
\;\dif \xi,
\end{equation}
for all $\widehat{\mathcal{A}}_i \in \mathcal{B}$ and $\mathcal{A}_i
\in C^\infty (M \times \Sigma, \mathbb{R})$. Such a correspondence
deforms the product of functions in $C^\infty(\Sigma,\mathbb{R})$
through the Moyal $\star$-product, which is defined by
\begin{equation}
\mathcal{F}_i \star \mathcal{F}_j := \mathcal{F}_i \exp \left(
\frac{\im \hbar}{2} \overleftrightarrow{\mathcal{P}} \right)
\mathcal{F}_j,       \nonumber
\end{equation}
where $\mathcal{F}_j = \mathcal{F}_j (\vec{x}, p, q; \hbar) \in
C^\infty(M \times \Sigma,\mathbb{R})$ and the operator
$\overleftrightarrow{\cal P}$ is given by
\begin{equation}
\overleftrightarrow{\cal P} :=
\frac{\overleftarrow{\partial}}{\partial q}
\frac{\overrightarrow{\partial}}{\partial p} -
\frac{\overleftarrow{\partial}}{\partial p}
\frac{\overrightarrow{\partial}}{\partial q}.    \nonumber
\end{equation}
At the same time, the Lie bracket between operators changes to the
Moyal bracket $\{ \cdot, \cdot\}_M$ between functions as follows:
\begin{equation}
\mathcal{W}^{-1} \left( \frac{1}{\im \hbar}
[\widehat{\mathcal{F}}_i, \widehat{\mathcal{F}}_j] \right) =
\frac{1}{\im \hbar} (\mathcal{F}_i \star \mathcal{F}_j -
\mathcal{F}_j \star \mathcal{F}_i) := \{\mathcal{F}_i, \mathcal{F}_j \}_M.
\end{equation}
As we said before, by taking the limit $\hbar \to 0$ we recover the
usual product between functions and the Poisson bracket,
respectively.

At this point, the deformation of equations (\ref{LocEqs}) can be
carry out as follows. First, let us consider the action
(\ref{action}), by promoting $A$ and $\phi$ to operator
$\mathfrak{su}(2)$-valued forms $\widehat{A}$ and $\widehat{\phi}$,
the action looks like
\begin{eqnarray}
S_q &=& -\int \dif^4 x \sqrt{g}\; \mathrm{Tr} \left[\frac{1}{2}
\widehat{F}_{\mu \nu} \widehat{F}^{\mu \nu} + \widehat{D}_\mu
\widehat{\phi}_\nu \widehat{D}^\mu \widehat{\phi}^\nu +
\widehat{R}_{\mu \nu} \widehat{\phi}^\mu \widehat{\phi}^\nu +
\frac{1}{2}
[\widehat{\phi}_\mu, \widehat{\phi}_\nu] [\widehat{\phi}^\mu, \widehat{\phi}^\nu] \right] \nonumber \\
& & + \frac{t - t^{-1}}{t + t^{-1}} \int_M \mathrm{Tr} \widehat{F}
\wedge \widehat{F} \label{op-action}
\end{eqnarray}

\noindent where the covariant derivative operator $\widehat{D}$ acting on the operator $\widehat{\phi}$ is given by $\widehat{D}\widehat{\phi} = \partial \widehat{\phi} + [\widehat{A},\widehat{\phi}]$. Even though we are writing the operator of the Ricci tensor $R_{\mu \nu}$, since its appearance is through the covariant derivative of the metric, $\dif = \nabla + A$, it can be treated as a function and not as an operator (in fact, it is proportional to the identity operator). Using by definition that
\begin{equation}
\mathrm{Tr} \widehat{(\cdot)} = 2\pi \hbar \sum_n \langle \psi_n |
\widehat{(\cdot)} | \psi_n \rangle,
\end{equation}
which is the sum of the diagonal elements with respect to the basis, and considering the previous setting, we incorporate these facts into the action (\ref{op-action}) with the promoted fields, becoming
\begin{eqnarray}
S_q &=& - 2\pi \hbar\sum_n \int \dif^4 x \sqrt{g}\; \left\langle
\psi_n \Big| \frac{1}{2(i \hbar)^2} \widehat{\mathcal{F}}_{\mu \nu}
\widehat{\mathcal{F}}^{\mu \nu} + \frac{1}{(i \hbar)^2}
\widehat{D}_\mu \widehat{\Phi}_\nu \widehat{D}^\mu
\widehat{\Phi}^\nu
+ \frac{1}{(i\hbar)^2} R_{\mu \nu} \widehat{\Phi}^\mu \widehat{\Phi}^\nu \right. \nonumber \\
&+& \left. \frac{1}{2(i \hbar)^2} \frac{1}{i \hbar}
[\widehat{\Phi}_\mu, \widehat{\Phi}_\nu] \frac{1}{i \hbar}
[\widehat{\Phi}^\mu, \widehat{\Phi}^\nu] \Big| \psi_n \right\rangle +  2\pi \hbar \frac{t - t^{-1}}{t + t^{-1}} \sum_n \int_M
\left\langle \psi_n \Big| \frac{1}{(i \hbar)^2}
\widehat{\mathcal{F}} \wedge \widehat{\mathcal{F}} \Big| \psi_n
\right\rangle    \nonumber  %\label{OP-action}    I
\end{eqnarray}
and hence, making a straight use of the Weyl correspondence
\cite{Zachos:2001ux}, the deformed action takes the form
\begin{eqnarray}
S_M &=& - 2\pi \hbar\int \dif^4 x\; \dif p\; \dif q \sqrt{g}\;
\frac{1}{\hbar^2} \left[\frac{1}{2} \mathcal{F}_{\mu \nu} \star
\mathcal{F}^{\mu \nu} + \mathcal{D}_{M\mu} \Phi_\nu \star
\mathcal{D}_M^\mu \Phi^\nu +
R_{\mu \nu} \Phi^\mu \star \Phi^\nu \right. \nonumber \\
& & + \left. \frac{1}{2} \{\Phi_\mu, \Phi_\nu\}_M \star \{ \Phi^\mu,
\Phi^\nu \}_M \right] + 2\pi \hbar\frac{t - t^{-1}}{t + t^{-1}}
\int_{M \times \Sigma} \dif p\; \dif q\; \frac{1}{\hbar^2}
\mathcal{F} \stackrel{\star}{\wedge} \mathcal{F}.
\label{def-action}
\end{eqnarray}

\noindent We have defined the \textit{Moyal covariant derivative} $\mathcal{D}_M = \dif + \{\mathcal{A}, \cdot \}_M$, where the action on the fields equals to\footnote{Here the derivatives are taken with respect to the coordinates of the manifold $M$.}  $\mathcal{D}_M\Phi = \dif\Phi + \{\mathcal{A},\Phi \}_M$  and the deformed field strength is given by
$\mathcal{F}_{\mu \nu} = \partial_\mu \mathcal{A}_\nu - \partial_\nu\mathcal{A}_\mu - \{ \mathcal{A}_\mu, \mathcal{A}_\nu \}_M.$ Note that the deformation process just acts on the functions defining the differential forms and not over the alternating tensors i.e. the deformation is only in the coordinates $p$ and $q$ of the symplectic surface $\Sigma$. Thus, following \cite{Strachan:1996gx} we define the \textit{Moyal wedge product}, denoted by $\stackon{$\wedge$}{$\star$}$, as
\begin{equation}
\omega \stackon{$\wedge$}{$\star$} \eta = \frac{1}{p! q!} \omega_{[i_1 \dots i_p} \star \eta_{j_1\dots j_q]} \dif x^{i_1} \wedge \dots \wedge \dif x^{j_q},
\end{equation}
for two $p$- and $q$-forms $\omega$ and $\eta$. These forms are defined by
\begin{equation}
\omega = \omega_{i_1 \dots i_p} (\vec{x}, p, q; \hbar) \dif x^{i_1}
\wedge \cdots \wedge \dif x^{i_p} \in \Omega^p(M, C^\infty (M \times
\Sigma, \mathbb{R})).
\end{equation}

At this point a natural question is: How the localization equations
look like under this deformation? It is straightforward to obtain
the action (\ref{def-action}) from this form
\begin{equation}
S = - 2\pi \hbar \int_{M \times \Sigma} \dif^4 x\; \dif p\; \dif q
\sqrt{g}\; \frac{1}{\hbar^2} \left[ \frac{t^{-1}}{t + t^{-1}}
\mathcal{V}_{_M \mu \nu}^+ \star \mathcal{V}_M^{+\mu \nu} +
\frac{t}{t + t^{-1}} \mathcal{V}_{M \mu \nu}^{-} \star
\mathcal{V}_M^{-\mu \nu} + \mathcal{V}_M^0 \star \mathcal{V}_M^0
\right]       \nonumber
\end{equation}
from where we obtain the Moyal localization equations
\begin{equation}
\mathcal{V}_M^+ = \mathcal{V}_M^{-} = \mathcal{V}_M^0 = 0,
\label{DefLocEqs}
\end{equation}
for
\begin{eqnarray}
\mathcal{V}_M^+ &=& (\mathcal{F} - \Phi \stackon{$\wedge$}{$\star$} \Phi + t \dif_\mathcal{A} \Phi )^+, \nonumber \\
\mathcal{V}_M^- &=& (\mathcal{F} - \Phi \stackon{$\wedge$}{$\star$} \Phi - t^{-1} \dif_\mathcal{A} \Phi )^-, \nonumber \\
\mathcal{V}_M^0 &=& \mathcal{D}_{M \mu} \Phi^\mu. \label{DefnonSW}
\end{eqnarray}

\indent As before, taking $t = 1$ and considering the definitions of (anti-)self-dual 2-forms in four dimensions, (\ref{DefLocEqs}) can be rewritten as
$$
\mathcal{F} - \Phi \stackon{$\wedge$}{$\star$} \Phi + *
\dif_\mathcal{A} \Phi = 0,
$$
\begin{equation}
\dif_\mathcal{A} * \Phi =0.
\end{equation}
These equations constitutes the Moyal deformation of Kapustin-Witten equations.

%%%%%%%%%%%%%%%%%%%%%%%%%%%%%%%%%%%%%%

\subsection{Looking for solutions}

\noindent Having discussed the general framework of the deformation quantization procedure, we are in position to find solutions to the deformed localization equations presented previously. Let $t_i$, $i = 1, \dots, 3$, be the generators of the Lie algebra $\mathfrak{su}(2)$ in an anti-hermitian representation and denote by $\widehat{\chi}_i$ the corresponding
$\mathfrak{su}(2)$ operators. Then we have the correspondence
\begin{eqnarray}
t_1 &\to& \widehat{\chi}_1 := i \beta \widehat{q} + \frac{1}{2 \hbar} (\widehat{q}^2 - \widehat{1}) \widehat{p}, \nonumber \\
t_2 &\to& \widehat{\chi}_2 := - \beta \widehat{q} + \frac{i}{2 \hbar} (\widehat{q}^2 + \widehat{1}) \widehat{p}, \nonumber \\
t_3 &\to& \widehat{\chi}_3 := i \beta \widehat{1} - \frac{1}{\hbar} \widehat{q} \widehat{p}, \nonumber
\end{eqnarray}
between the generators of $\mathfrak{su}(2)$ and $\mathfrak{su}(2)$-valued operators which act on some Hilbert space. The parameter $\beta$ is due to some choice in the ordering between $\widehat{q}$ and $\widehat{p}$. Under the Weyl isomorphism these operators correspond to functions defined on the symplectic surface $\Sigma$ using the formula
\begin{equation}
\chi_i (\vec{x}, p, q; \hbar) := \int_{-\infty}^{\infty} \left\langle q - \frac{1}{2} \xi \Big| \widehat{\chi}_i \Big| q + \frac{1}{2} \xi \right\rangle \exp \left( \frac{i}{\hbar} \xi \cdot p \right) \dif \xi, \nonumber
\end{equation}
from where we get the corresponding functions
\begin{equation}
\chi_1 (p, q; \hbar) = i \left( \beta - \frac{1}{2} \right) q -
\frac{1}{2 \hbar} (q^2 - 1) p, \label{ChIoNe}
\end{equation}
\begin{equation}
\chi_2 (p, q; \hbar) = -\left( \beta - \frac{1}{2} \right) q -
\frac{i}{2 \hbar} (q^2 + 1) p, \label{ChItWo}
\end{equation}
\begin{equation}
\chi_3 (p, q; \hbar) = -i \left( \beta - \frac{1}{2} \right) +
\frac{1}{\hbar} q p. \label{ChItHrEe}
\end{equation}

\noindent With respect to the Moyal bracket these functions satisfy the $\mathfrak{su}(2)$ algebra relations
\begin{equation}
\{ \chi_1, \chi_2 \}_M = -\frac{1}{\im \hbar} \chi_3, \qquad (\mbox{plus cyclic permutations}).
\end{equation}

The solutions we found earlier are promoted to their corresponding operators and by applying the Weyl correspondence, they look like
\begin{eqnarray}
A_1 + i A_2 = i (\partial_1 + i \partial_2) v t_3 &\rightarrow& \mathcal{A}_1 + i \mathcal{A}_2 = -\hbar (\partial_1 + i \partial_2) v \chi_3, \\
\phi_0 = - \partial_3 v t_3 &\rightarrow& \Phi_0 = -i \hbar \partial_3 v \chi_3, \\
\phi = z^\mathfrak{r} e^v (t_2 - i t_1) &\rightarrow& \mathcal{P} = i \hbar z^\mathfrak{r} e^v (\chi_2 - i \chi_1),
\end{eqnarray}
where the field strength is given by
\begin{equation}
\mathcal{F}_{12} = i \hbar (\partial_1^2 + \partial_2^2) v \chi_3,
\end{equation}
and where $v$ satisfy the differential equation (\ref{Eqforv}). This set of functions fulfill the corresponding deformed system of conditions
\begin{eqnarray}
\{ \mathcal{D}_{Mi}, \mathcal{D}_{Mj}\}_M &=& 0, \qquad i, j = 1, \dots, 3, \\
\sum_{i = 1}^3 \{ \mathcal{D}_{Mi}, \mathcal{D}_{Mi}^\dagger \}_M &=& 0,
\end{eqnarray}

\noindent for the components
\begin{eqnarray}
\mathcal{D}_{M1} &=& \frac{\partial}{\partial x^1} + i \frac{\partial}{\partial x^2} + \{ \mathcal{A}_1 + i \mathcal{A}_2, \cdot\; \}_M, \nonumber \\
\mathcal{D}_{M2} &=& \frac{\partial}{\partial x^3} + \{ \mathcal{A}_3 - i \Phi_0, \cdot\; \}_M, \nonumber \\
\mathcal{D}_{M3} &=& \{ \Phi_1 - i \Phi_2, \cdot\; \}_M.
\end{eqnarray}

%%%%%%%%%%%%%%%%%%%%%%%%%%%%%%%%%%%%%%%%%%%%%%%%%%%
%%%%%%%%%%%%%%%%%%%%%%%%%%%%%%%%%%%%%%%%%%%%%%%%%%%

\section{Non-abelian Seiberg-Witten equations}

\noindent As we have discussed before, on the localization equations
(\ref{LocEqs}) we can impose a further conditions to these equations
asking that, for any $t$, $F - \phi \wedge \phi$ be self-dual and $D
\phi$ be anti-selfdual. By imposing these conditions we get the so
called \textit{non-abelian Seiberg-Witten equations}
\cite{Dunajski:2011sx}
\begin{eqnarray}
\mathcal{V}^+ &=& (F - \phi \wedge \phi )^+ = 0, \nonumber \\
\mathcal{V}^{-} &=& (D\phi)^- = 0, \nonumber \\
\mathcal{V}^0 &=& D^* \phi = 0. \label{nonAbSW}
\end{eqnarray}

These equations has a striking resemblance to the Hitchin's equations, which are defined for one complex dimension manifolds. In fact, Ward in \cite{Ward:2016ygr} generalizes these  equations to higher dimensions. Similar to the case of (\ref{LocEqs}), the action leading to (\ref{nonAbSW}), up to boundary terms, is written as
\begin{eqnarray}
S &=& -\int \dif^4 x \sqrt{g}\; \mathrm{Tr} \left[ \mathcal{V}_{\mu \nu}^+ \mathcal{V}^{+ \mu \nu} + \mathcal{V}_{\mu \nu}^- \mathcal{V}^{- \mu \nu} + \mathcal{V}_0^2 \right] \nonumber \\
&=& -\int \dif^4 x \sqrt{g}\; \mathrm{Tr} \left[\frac{1}{2} F_{\mu
\nu}F^{\mu \nu} + D_\mu \phi_\nu D^\mu \phi^\nu + R_{\mu \nu}
\phi^\mu \phi^\nu + \frac{1}{2} [\phi_\mu, \phi_\nu] [\phi^\mu,
\phi^\nu] \right].  \nonumber
\end{eqnarray}

As we did with (\ref{LocEqs}), we perform a deformation of these equations via the WWMG formalism; provided that we have nearly the same type of equations previously studied in the above section, such deformation follows a similar procedure. In fact, after promoting each field to an operator valued one and applying the Weyl correspondence to (\ref{nonAbSW}), we have
\begin{eqnarray}
\mathcal{V}_M^+ &=& (\mathcal{F} - \Phi \stackon{$\wedge$}{$\star$} \Phi )^+ = 0, \nonumber \\
\mathcal{V}_M^{-} &=& (\mathcal{D}_M \Phi)^- = 0, \nonumber \\
\mathcal{V}^0 &=& \mathcal{D}_{M\mu} \Phi^\mu = 0,
\label{DnonAbSW}
\end{eqnarray}

\noindent which are the integrable deformation analogous to (\ref{DefnonSW}). At the same time, the deformation for the corresponding action is
\begin{eqnarray}
S_M &=& \int \dif^4 x\; \dif p\; \dif q \sqrt{g}\; \frac{1}{\hbar^2} \left[\frac{1}{2}
\mathcal{F}_{\mu \nu} \star \mathcal{F}^{\mu \nu} + \mathcal{D}_{M\mu} \Phi_\nu \star \mathcal{D}_M^\mu \Phi^\nu + R_{\mu \nu} \Phi^\mu \star \Phi^\nu \right. \nonumber \\
& & + \left. \frac{1}{2} \{\Phi_\mu, \Phi_\nu\}_M \star \{ \Phi^\mu,
\Phi^\nu \}_M \right], \nonumber
\end{eqnarray}
which has exactly the form of (\ref{def-action}). It is interesting
to note that (\ref{nonAbSW}) have solutions close in spirit for the
generalized Kapustin-Witten equations. As reported in
\cite{Dunajski:2011sx}, we have solutions to the original
non-abelian Seiberg-Witten equations for different backgrounds. we
will review this in the next subsection.

%%%%%%%%%%%%%%%%%%%%%%%%%%%%%%%%%%%%%%%%%%%%%%

\subsection{Some solutions}

\noindent In \cite{Dunajski:2011sx} Dunajski and Hoegner found some solutions to the non-abelian Seiberg-Witten equations relying in some ansatz for functions satisfying a certain
set of differential equations. From the self-duality equations defined on a $\mathrm{Spin}(7)$-holonomy manifold $M_8$ for some group $G$, splitting this manifold into two four-dimensional hyper-K\"ahler manifolds $M'_4 \times M_4$ with a proper choice of the components of the connection 1-form and a given dimensional reduction suggested by this splitting, the authors obtain (\ref{nonAbSW}) with gauge group $\mathrm{SU}(2)$. The appearance of such particular group comes from the fact that with this choice of
hyper-K\"ahler manifolds a holonomy reduction is induced, provided that $\mathrm{SU}(2) \times \mathrm{SU}(2)$ is a proper subgroup of $\mathrm{Spin}(7)$. Details of how all this is done can be referred to the original paper \cite{Dunajski:2011sx}. Here, we are mainly concerned with the solutions they found in doing such selection.

Remember that we have chosen $t_a$ as our generators of the Lie algebra $\mathfrak{su}(2)$ in an anti-hermitian representation. For all the fields involved there is no dependence on the coordinates of $M'_4$. Let $A$ and $\phi$ be again the connection 1-form and a 1-form of scalar fields on $M_4$, respectively, that are $\mathfrak{su}(2)$-valued. Also on $M_4$, let $e^a$ be our vierbein, and define the two-forms
\begin{equation}
\psi_i^+ = e^0 \wedge e^i + \frac{1}{2} \varepsilon_{\; jk}^i e^j
\wedge e^k,  \nonumber
\end{equation}
which are self-dual with respect to the Hodge operator of $M_4$. Let $G$ and $H$ be scalar functions on $M_4$, then the ansatz for $A$ and
$\phi$ proposed in \cite{Dunajski:2011sx} reads
\begin{eqnarray}
A = * \left( \sum_i t_i \psi_i^+ \wedge \dif G \right) = \sum_i t_i * (\psi_i^+ \wedge \dif G), \nonumber \\
\phi = * \left( \sum_i t_i \psi_i^+ \wedge \dif H \right) = \sum_i
t_i * (\psi_i^+ \wedge \dif H).
\label{ansatz}
\end{eqnarray}
Here, we intentionally separate the Lie algebra generator from the two-form in order to make explicit how the WWMG correspondence will be implemented.

In order to satisfy (\ref{nonAbSW}) with the proposed $A$ and $\phi$, the scalars functions $G$ and $H$ must be solutions to the next set of differential equations:
\begin{eqnarray}
0 &=& \Box G + |\nabla G|^2 - |\nabla H|^2, \label{eq1} \\
0 &=& (\varepsilon_{ea}^{\;\;\; bc} C_{\;\; bc}^{a} \sigma^{ed} -
\sigma^{ab} C_{\;\; ab}^{d}) \nabla_d G,
\label{eq2} \\
0 &=& \tilde{\sigma}_{ac} \sigma_{\;\; b}^{c} (\nabla^a \nabla^b H - 2 \nabla^a G \nabla^b H),
\label{eq3} \\
0 &=& \sigma_{ab} (\nabla^a \nabla^b H - 2 \nabla^a G \nabla^b H),
\label{eq4}
\end{eqnarray}
with $\Box$ and $\nabla$ differential operators on $M_4$; $C_{\;\;bc}^{a}$ are the structure constants defined by $\dif e^a = C^{a}_{\;\; bc} e^b \wedge e^c$.

At this point, we can obtain the functions $G$ and $H$ explicitly for different backgrounds and at the same time we can write down its deformation via the WWMG formalism.

%%%%%%%%%%%%%%%%%%%%%%%%%%%%%%%%%%%%%%%%%%%%%

\subsubsection{A simple case, flat background}

Let $M_4 = \mathbb{R}^4$. In this case we have that $e^i = \dif
x^i$, thus $C^{a}_{\;\; bc} = 0$. The connection has the explicit
form
\begin{eqnarray}
A &=& t_1 \left(\partial_0 G \dif x^1 - \partial_1 G \dif x^0 + \partial_2 G \dif x^3 - \partial_3 G \dif x^2 \right)\nonumber \\
&+& t_2 \left(\partial_0 G \dif x^2 - \partial_2 G \dif x^0 + \partial_3 G \dif x^1 - \partial_1 G \dif x^3 \right) \nonumber \\
&+& t_3 \left(\partial_0 G \dif x^3 - \partial_3 G \dif x^0 +
\partial_1 G \dif x^2 - \partial_2 G \dif x^1 \right).
\end{eqnarray}
And a similar expression for $\phi$. When these 1-forms are inserted in the set of equations
(\ref{eq1})-(\ref{eq4}) with the given structure constants, the corresponding solutions to the functions $G$ and $H$ read
\begin{equation}
G = -\frac{1}{2} \ln |x^3|, \qquad H = \frac{\sqrt{3}}{2} \ln |x^3|.  \nonumber
\end{equation}

In order to apply the WWMG correspondence, we promote as before $A$ and $\phi$ to operator valued quantities, where the $\mathfrak{su}(2)$ generators are now $\mathfrak{su}(2)$-valued operators, that is $t_a \to \widehat{t}_a$, satisfying the $\mathfrak{su}(2)$ algebra relations. Thus, the operator form of the connection 1-form is
\begin{equation}
\widehat{A} = \frac{\partial G}{\partial x^3} \left[ - \widehat{t}_3
\dif x^0 + \widehat{t}_2 \dif x^1 - \widehat{t}_1 \dif x^2 \right]  \nonumber
\end{equation}

\noindent with a similar expression for $\phi$. Hence, by applying the Weyl isomorphism, we obtain the deformed 1-form
\begin{equation}
\mathcal{A} = \frac{\partial G}{\partial x^3} \left[\chi_3 (p, q; \hbar)
\dif x^0 + \chi_2 (p, q; \hbar) \dif x^1 + \chi_1 (p, q; \hbar) \dif x^2 \right],
\end{equation}

\noindent where the functions $\chi_i$ are functions of the symplectic surface $\Sigma$ with expressions given by the equations (\ref{ChIoNe})-(\ref{ChItHrEe}). These functions satisfy the Moyal $\mathfrak{su}(2)$-relations
\begin{equation}
\{ \chi_1, \chi_2 \}_M = -\frac{1}{\im \hbar} \chi_3, \qquad (\mbox{plus cyclic permutations}),
\end{equation}

\noindent and we have a similar expression for $\Phi (\vec{x}, p, q;\hbar)$.

%%%%%%%%%%%%%%%%%%%%%%%%%%%%%%%%%%%%%%%%%%%%

\subsubsection{Curved backgrounds}

\noindent As the previous example shows, the Weyl correspondence acts just on the Lie algebra generators of the 1-forms $A$ and
$\phi$ in (\ref{ansatz}). Hence, the $\mathfrak{su}(2)$ operator-valued 1-forms are
\begin{eqnarray}
\widehat{A} = * \left( \sum_i \widehat{t}_i \psi_i^+ \wedge \dif G \right) = \sum_i \widehat{t}_i * (\psi_i^+ \wedge \dif G), \nonumber \\
\widehat{\phi} = * \left( \sum_i \widehat{t}_i \psi_i^+ \wedge
\dif H \right) = \sum_i \widehat{t}_i * (\psi_i^+ \wedge \dif H).
\label{OPansatz}
\end{eqnarray}

\noindent Applying the Weyl correspondence to these 1-forms gives the expression
\begin{eqnarray}
\mathcal{A} = * \left( \sum_i \chi_i \psi_i^+ \wedge \dif G \right) = \sum_i \chi_i *
(\psi_i^+ \wedge \dif G), \nonumber \\
\Phi = * \left( \sum_i \chi_i \psi_i^+ \wedge \dif H \right) =
\sum_i \chi_i * (\psi_i^+ \wedge \dif H), \label{WWansatz}
\end{eqnarray}

\noindent with the functions of the phase space $\chi_i$ previously defined.\\
\indent Now, let us consider as another example the hyper-K\"ahler metric
\begin{equation}
\mathrm{g}_4 = V \big[(\dif x^1)^2 + (\dif x^2)^2 + (\dif
x^3)^2\big] + V^{-1} (\dif x^0 + \alpha)^2,     \nonumber
\end{equation}

\noindent where the function $V$ and the 1-form $\alpha$ depend on $x^i$, $i = 1, \dots, 3$, and are related by $*_3 \dif V = -\dif \alpha$, with $*_3$ the Hodge operator in $\mathbb{R}^3$. A solution for the corresponding equations is given by
\begin{equation}
G = -\frac{3}{4} \ln x^3 + \frac{1}{4} \ln 21 - \ln 2, \qquad H = -\frac{\sqrt{21}}{3} G.  \nonumber
\end{equation}

\noindent By using the Weyl correspondence once again, we obtain that the next forms are solutions to the non-abelian Seiberg-Witten-Moyal
equations
$$
\mathcal{A} = \frac{3}{4} (\sigma_2 \otimes \chi_1 - \sigma_1 \otimes \chi_2 + \sigma_0 \otimes \chi_3), \qquad \Phi = -\frac{\sqrt{21}}{3} \mathcal{A}, \\
$$
$$
\mathcal{F} = \left(\frac{9}{16} \sigma_0 \wedge \sigma_1 + \frac{3}{4} \sigma_2 \wedge \sigma_3 \right) \otimes \chi_1
$$
\begin{equation}
 + \left(\frac{9}{16} \sigma_0 \wedge \sigma_2 - \frac{3}{4} \sigma_1 \wedge \sigma_3 \right) \otimes \chi_2 + \left(\frac{3}{2} \sigma_0 \wedge \sigma_3 - \frac{3}{16} \sigma_1 \wedge \sigma_2 \right) \otimes \chi_3, \nonumber
\\
\end{equation}
where the 1-forms $\sigma$'s are defined by
\begin{equation}
\sigma_0 = (x^3)^{-2} (\dif x^0 + x^2 \dif x^1), \qquad \sigma_1 =
(x^3)^{-1} \dif x^1, \qquad \sigma_2 = (x^3)^{-1} \dif x^2, \qquad
\sigma_3 = \frac{\dif x^3}{x^3}.    \nonumber
\end{equation}

In principle, similar solutions like the ones presented in this
section can be obtained when the WWMG formalism is applied to the
integral generalization to higher dimensions of the Hitchin
equations performed by Ward in \cite{Ward:2016ygr}, something that
is under development at the present. Generalizations to the Hitchin
equations to higher dimensions can be found also in \cite{Simpson}.
In this regard, Ward make emphasis on generalizations which are at
the same time integrable; he gave such generalization for dimensions
equal to $2k$. The model for him is the octonionic self-dual
equations defined on a eight-dimensional $\mathrm{Spin}(7)$-holonomy
manifold, which are also the case for Dunajski and Hoegner as we
already study, and for Cherkis in \cite{Cherkis:2014xua}, when he
deduce the Haydys-Witten equations from a dimensional reduction of
the mentioned octonionic system.

%%%%%%%%%%%%%%%%%%%%%%%%
%%%%%%%%%%%%%%%%%%%%%%%%
%%%%%%%%%%%%%%%%%%%%%%%%

\section{Final Comments}

Integrability is an aspect of a broad class of system of equations
appearing in many contexts of physical and mathematical interest
alike. In general, there are conditions to decide when a given
system is integrable, but there is no general rule that apply to all
systems. Such is the case, for example, of the integrability of the
full Yang-Mills equations, which is still an open problem.

In this paper we performed an integrable deformation of the so
called Kapustin-Witten equations and the non-abelian Seiberg-Witten
equations via the WWMG formalism. It is already known that these set
of equations are integrable, thus spoiling nothing when we perform
such deformation; also, solutions to these equations together with
its corresponding deformations were presented. The possibility of
carrying out such deformation in Hitchin's systems on $\mathbb{R}^2$
\cite{WardTwo} and in dimensions greater than two
\cite{Ward:2016ygr} are being explored and will be reported in a
future communication.

%%%%%%%%%%%%%%%%%%%%%%%%%%%%%%%%%%%%%%%%%%%%%%%%%%%%%%%%%%%%%%%%%%%%%%%%%%
 \vspace{.5cm}
\centerline{\bf Acknowledgments} \vspace{.5cm} The work of H. G-C. was partially supported by SNI-M\'exico and by the CONACyT research grant: 128761. A. Mart\'inez-Merino would like to thank to Divisi\'on de Ciencias e Ingenier\'ia for the hospitality where this work initially began.

%%%%%%%%%%%%%%%%%%%%%%%%%%%%%%%%%%%%%%%%%%%%%%%%%%%%%%%%%%%%%%%%

\appendix
\section{On the Hitchin equations}

The Hitchin equations first appear in a classical article by Hitchin \cite{Hitchin:1987} as a two dimensional reduction of the selfdual Yang-Mills equations in four dimensions. In geometric terms, such equations can be defined for a $G$-principal bundle over a riemannian 2-manifold $M$ and are given by:
\begin{equation}
F+[\Phi,\Phi^{*}]=0\,,\quad\quad \bar\partial_{A}\Phi=0\,. \label{Hitchin eqs.}
\end{equation}
Here $A$ is a connection on the bundle with $F$ its curvature, $\Phi$ is a certain $\mathfrak{g}$-valued holomorphic 1-form with adjoint $\Phi^{*}$ and $\bar\partial_{A}=D_{1}+iD_{2}$. These equations are close related with the Kapustin-Witten equations and in the literature they usually appear in different forms. For instance, in \cite{Kapustin:2006pk} the eqs. (\ref{Hitchin eqs.}) are written as:
\begin{equation}
F-\Phi\wedge\Phi = 0\,,\quad\quad D\Phi=0\,, \quad\quad D*\Phi=0\,, \label{Hitchin eqs. 2}
\end{equation}
where $D$ is the covariant derivative of $A$ with gauge field $F=dA+ A\wedge A$ and $\Phi$ is again a $\mathfrak{g}$-valued 1-form. The systems of equations (\ref{Hitchin eqs.}) and (\ref{Hitchin eqs. 2}) are indeed the same, the difference arises because the objects are arranged in a different way. In (\ref{Hitchin eqs.}) the form
$\Phi=\frac{1}{2}\phi\,dz$, where $\phi=\phi_{1}-i\phi_{2}$ and
$z=x^{1}+ix^{2}$. Instead, in (\ref{Hitchin eqs. 2}) the form $\Phi=\phi_{1}dx^{1} + \phi_{2}dx^{2}$. In both cases $\phi_{1}$ and $\phi_{2}$ are the Higgs fields induced by the dimensional reduction procedure and the equivalence is almost evident if we notice that $\phi_{1}$ and $\phi_{2}$ are antihermitian and we write the
$\mathfrak{g}$-valued form in (\ref{Hitchin eqs.}) as $\Phi_{c}=\frac{1}{2}(\phi_{1}-i\phi_{2})dz$, where the subscript $c$ remember us that it is a complex Higgs field. In fact, using
this notation $\Phi^{*}_{c}=-\frac{1}{2}(\phi_{1}+i\phi_{2})d\bar z$ and we get
\begin{equation}
[\Phi_{c},\Phi^{*}_{c}] =  -\frac{i}{2}[\phi_{1},\phi_{2}]dz\wedge
d\bar z = -[\phi_{1},\phi_{2}]dx^{1}\wedge dx^{2} = -\Phi\wedge\Phi,
\nonumber
\end{equation}
which shows that the first eq. in (\ref{Hitchin eqs.}) corresponds to the first equation in (\ref{Hitchin eqs. 2}). Now
\begin{equation}
\bar\partial_{A}\Phi_{c} = \frac{1}{2}(D_{1}+iD_{2})(\phi_{1}-i\phi_{2})dz = \frac{1}{2}[D_{1}\phi_{1} +D_{2}\phi_{2} +i(D_{2}\phi_{1} - D_{1}\phi_{2})]dz \nonumber
\end{equation}
and hence the second eq. in (\ref{Hitchin eqs.}) is equivalent to $D_{1}\phi_{1} +D_{2}\phi_{2}=0$ and $D_{2}\phi_{1} - D_{1}\phi_{2}=0$, but these are precisely the expressions in
components of $D*\Phi=0$ and $D\Phi=0$.

%%%%%%%%%%%%%%%%%%%%%%%%%%%%%%%%%%%%%%%%%%%%%%%%%%%%%%%%%%%%%%%%%%%%%%%%%%%%%%%%%%%%%%
%%%%%%%%%%%%%%%%%%%%%%%%%%%%%%%%%%%%%%%%%%%%%%%%%%%%%%%%%%%%%%%%%%%%%%%%%%%%%%%%%%%%%%
%%%%%%%%%%%%%%%%%%%%%%%%%%%%%%%%%%%%%%%%%%%%%%%%%%%%%%%%%%%%%%%%%%%%%%%%%%%%%%%%%%%%%%

%\vskip 2truecm
%%%%%%%%%%%%%%%%%%%%%%%%%%%%%%%%%%%%%%%%%%%%%%%%%%%%%%%%%%%%%%%%%%%%%%%%%%

\end{document}